# A breakthrough toward wafer-size epitaxial graphene transfer

A. Hashimoto[1], H. Terasaki[1], K. Morita[3], H. Hibino[2]  &  S. Tanaka[3]

**The formation of graphene on any desirable substrate is extremely essential for the successful replacement of Si with graphene in all technological applications in the beyond-CMOS era. Recently, we observed that a Ti layer formed on epitaxial graphene by electron-beam evaporation has the sufficiently large area of contact with the surface of epitaxial graphene for the exfoliation process to take place. We also observed that the exfoliated FLG on the Ti layer can be easily transferred onto a $SiO_2$/Si substrate. In this paper, we have proposed a new technique for transferring graphene having a large area of several square mm from the graphitized vicinal SiC substrate. Raman scattering spectroscopy and low-energy electron microscopy analysis have revealed that we can transfer mono-layer and bi-layer graphene onto the $SiO_2$/Si substrate by using the proposed transfer process. The initial size of epitaxial graphene formed on the SiC substrate is the only limitation of the new transfer process. The transfer process is expected to become an extremely important technology that will mark the beginning of a new era in the field of graphene electronics.**

1. Graduate School of Electrical & Electronics Engineering, University of Fukui, Bunkyo 3-9-1, Fukui 910-8507, Japan
2. NTT Basic Research Laboratories, NTT Corporation, Atsugi, Kanagawa 243-0198, Japan
3. Department of Applied Quantum Physics & Nuclear Engineering, Kyusyu University, 744 Motooka, Nishi-ku, Fukuoka 819-0395, Japan

Correspondence to: Akihiro Hashimoto [1], e-mail: hasimoto@fuee.fukui-u.ac.jp

**Introduction**

Graphene, a single mono-layer of graphite, has attracted considerable interest because of its exotic carrier transport properties at temperatures up to room temperature. [1-3] It has excellent potential to replace Si in technological applications in the beyond-CMOS era. For such applications, a technology that can facilitate the formation of few-layer graphene (FLG) on any desirable substrate is absolutely essential. Some conventional techniques such as the micromechanical exfoliation of isolated graphene (maximum areas: ~ 1 mm$^2$) from bulk graphite, are essentially unsuitable for the large-scale production of graphene-based devices. On the other hand, it is expected that the graphitization of SiC (0001) under the ultra-high-vacuum (UHV) condition or in an Ar atmosphere will produce FLG with a considerably larger size than that previously attainable. [4-7] However, it is a little difficult to directly use the epitaxial graphene formed on a SiC substrate in actual device applications, because the interesting properties of the graphene would be reduced by an ambiguous interfacial problem between graphene and the surface of the SiC substrate and the strong limitation by the usage of SiC substrates exists for the actual device applications. [7, 8] In order to avoid these problems, we need to transfer the FLG from the SiC surface to a desirable substrate such as SiO$_2$/Si. The key idea is to use the epitaxial graphene on the SiC substrate as a graphene seed layer in the process of transferring graphene onto a desirable surface. [9] Although the sample size in this paper was typically 8 mm $\times$ 8 mm, which was limited only by the annealing apparatus used in this study, it is expected that we can obtain the epitaxial graphene as large as the size of the SiC substrates. We can confirm the considerably high quality of epitaxial graphene on the Si-terminated SiC substrates by using Raman scattering spectroscopy, angular-resolved photo-emission spectroscopy (ARPES), and low-energy electron microscopy (LEEM). The electron mobility of epitaxial graphene on a Si-terminated SiC (0001) substrate has been characterized by several research groups. The value of electron mobility in the case of epitaxial graphene at room temperature typically runs into several thousands of cm$^2$/Vs. Although this value is not as high as the maximum value for HOPG graphene, some researchers have attributed the reduction in electron mobility to the interface between the FLG and the SiC substrate. [10] Therefore, epitaxial graphene described here can be used as a graphene seed layer in the process of transferring FLG onto another substrate: the new transfer process proposed in this paper is expected to establish a method for the formation of a large-area graphene/insulator structure on a technologically viable scale.

**Experimental**

The preparation of epitaxial graphene as a seed layer through the thermal decomposition of a vicinal SiC substrate has been carried out using an annealing process under the conditions of current heating and UHV. [11-14] It is well known that the thermal decomposition of an SiC substrate leads to the formation of large, uniform epitaxial graphene (thickness: a few mono-layer) on the Si-terminated (0001) vicinal SiC plane with a 4° incline in the [11-20] direction. The multi-layered and the rotationally disordered graphene layers have been produced on the C-terminated (000-1) face. [15, 16] In this study, the epitaxial graphene seed layers were synthesized on commercial, vicinal Si-terminated 6H-SiC (0001) substrates. Prior to graphene epitaxy, the sample was etched in a hydrogen ambience at 1430 ℃ at 760 Torr for 15 min in order to remove any damage caused by surface polishing and to form a step-ordered structure on the surface using a horizontal cold-wall reactor with an infrared-lamp-annealing system. Graphene epitaxy was carried out in an UHV chamber by using a direct heating system. The heating and cooling rates were 2-3 ℃ per second. The typical direct annealing time and temperature were 15 min and 1650 ℃, respectively. A detailed mechanism of the formation of uniform epitaxial graphene on the vicinal step-ordered Si-terminated SiC surface has been published in another related article. [14]

Figure 1 shows a photo-image of a typical epitaxial graphene (size: 4 mm × 8 mm) formed on a vicinal Si-terminated SiC substrate. The sample size of epitaxial graphene was only limited by the scale of the annealing apparatus used in the study. Therefore, the following results of the present study were not affected by any limitation of the sample size. A large epitaxial graphene seed layer can be completely transferred onto another substrate by using the proposed transfer process. The LEEM images and the phase images of the atomic force microscope (AFM) are also shown with the Raman spectrum in Fig. 1. In the LEEM image, the bi-layer region of epitaxial graphene was mainly observed but tri-layer region was also observed along the vicinal steps in ~10% of the surface area. Further, note that the LEEM image corresponded exactly to the contrast of the AFM phase image as shown in Fig.1. Therefore, we can estimate the thickness variation, and not the thickness itself, of epitaxial graphene: the transferred graphene could also be estimated by using the AFM phase image. The Raman spectrum, shown in Fig.1, indicated that the quality of epitaxial graphene as estimated from the narrow FWHM of the Raman peaks and a very small portion of the D-band intensity was rather excellent although the peaks of the G- and the G'-bands showed a considerable blue-shift because of the strong compressive stress from the SiC substrate. [17] ARPES also revealed that the electronic structure of the bi-layer graphene could be observed clearly.

A schematic representation of the proposed new transfer process is shown in Fig. 2. The transfer process consists of two stages named the first stage and the second stage. The pristine epitaxial graphene seed layer formed on the vicinal Si-terminated SiC (0001) surface was covered with a metal Ti layer formed at room temperature by the electron beam evaporation under the high vacuum condition of less than $10^{-6}$ Pa of background pressures in the first stage. The thickness of the metal Ti layer in the present study was more than 15 nm, and the metal Ti layer thinner than 15 nm was not effective in the exfoliation process. The surface morphology of the pristine epitaxial graphene was very smooth with an RMS value of less than 0.97 nm including the ordered steps, but an RMS value of the terraces is less than 0.12 nm; however, the morphology after the evaporation of the Ti metal layer became rather rough with an RMS value of more than 1.1 nm. Epoxy-type glue was coated on the surface of the metal Ti layer and an Al block of the force gauge system that was used for the quantitative exfoliation process. The other side of the SiC substrate was also directly fixed on another Al block by using the epoxy-type glue because the back-surface of the SiC substrate was unpolished and considerably rough. The binding force exerted by the van der Waals interactions between the FLG and the SiC surface during the exfoliation process of the bi-layer graphene was estimated by using a typical force curve and was found to be around ~ 0.8 MPa. Because the binding force for the exfoliation process of HOPG is typically ~ 0.2 MPa, the binding force between the FLG and the SiC surface in the epitaxial graphene seed layers was considerably stronger than that in HOPG, because of the large-area epitaxial stacking for the graphene on the SiC surface against the random rotational stacking for the HOPG. The strong binding force of around 30 MPa between the Ti surface or the SiC back surface and the Al blocks of the force gauge system were due to the geometrical anchor effect caused by the permeation of the glue into the rough surface: this effect was sufficiently strong for the exfoliation of the graphene from the graphitized SiC surface to occur. We found that the epitaxial graphene seed layer just above the buffer layer could be removed as discussed in the following section.

**Results and discussion**

Figure 3 shows the photo-images of the transferred graphene on the Ti layer and/or the SiC substrates after the first stage in the transfer process. Figure 3 (a) shows the schematic representations of the graphene on the Ti layer with the AFM phase image and the optical photo image. The Raman spectra from the graphene on the Ti layer after the first stage of the transfer process are also shown in Fig. 3(a). Figure 3 (b) shows the schematic representations of the remaining SiC surface after the first stage with the LEEM image and the LEED pattern. The Raman spectra from the sample are also shown in the Fig 3(b). In this study, we could transfer almost all of the bi-layer graphene (area: 4 mm×7 mm) from the pristine graphene seed layer onto the Ti layer, as shown in the AFM and the LEEM images in Fig. 3. From the LEEM measurements, we could also confirm that the SiC surface after the first stage in the transfer process held less monolayer graphene; that is, the graphene seed layer of the bi-layer graphene was exfoliated from the pristine surface of the SiC substrate onto the surface of the Ti layer. On the SiC surface after the exfoliation, we could observe a clear $6\sqrt{3} \times 6\sqrt{3}$ R30 reconstruction after the thermal annealing at 600 ℃ under UHV condition, as shown in the LEED images of Fig. 3. This means that the connection between the graphene layer of the epitaxial growth front and the buffer layer was weaker than the van der Waals connection of each graphene layer. We also observed thick graphene domains along the steps, as shown in LEEM image of Fig. 3(b);

these domains could be attributed to the thickness variation of pristine epitaxial graphene, as shown in Fig. 1. Microscopic Raman scattering spectroscopic measurements from the graphene layers transferred onto the surface of the Ti layer and from the remaining SiC surface were performed at room temperature in the back-scattering geometry. All observed Raman spectra from the graphene on the Ti layer showed clear G-band and G`-band modes, although there was no G`-band peak in the Raman spectra of the remaining SiC surface, as shown in Fig. 3. It has been reported that the width and the shape of the G`-band mode strongly depended on the layer thickness and the electronic structure because the G`-band is a result of the second-order resonant Raman process of the graphene. [18] Therefore, it is possible to analyze the band structure and the layer thickness of FLG by a Raman analysis. In this study, we focused on the studying the uniformity of the layer thickness and the electronic structure of the graphene transferred onto the Ti layer. The Raman spectra at several points in the graphene layer transferred onto the Ti layer strongly indicated that the transferred graphene was very uniform and the thickness corresponded exactly to the bi-layer graphene estimated by the analysis of the G`-band. The Raman results were consistent with the LEEM observation for the remaining SiC surface: that is, the transferred graphene on the Ti layer had a bi-layer thickness. The comparison of the Raman spectra between the pristine bi-layer graphene on the SiC substrate and the transferred graphene on the Ti layer from the epitaxial graphene seed layer showed that there were large red-shifts of the G- and G`-peaks in the case of the transferred graphene, as shown in Fig. 3. The Raman results indicated that the strong compressive stress on epitaxial graphene on the SiC substrate was released after the transfer onto the Ti layer. The other exfoliation experiments for the epitaxial graphene seed layer with tri-layer thickness were also successful, and the graphene layer transferred onto the Ti layer showed that the Raman spectra corresponded to the graphene layer that had tri-layer thickness.

The second-order Raman peak, G`-band, analysis, shown in Fig. 4 (a) indicated that the Raman spectrum from the graphene transferred onto the Ti layer could be decomposed into three sub-peaks corresponding to the second-order Raman transitions denoted by $P_{11}$, $P_{12}+P_{21}$, and $P_{22}$. The schematic representations of each second-order transition process are also shown in Fig. 4 (b).

The intensity ratio of each component in the second-order Raman spectra showed good agreement with the theoretical prediction, although the decomposition result of the G`-band from the micromechanical exfoliated bi-layer graphene from HOPG was not similar to the theoretical prediction by J. S. Park et al., even after including the excitation energy dependence, as shown in Fig. 4 (a). [19] This means that the graphene transferred onto the Ti layer had a considerably ideal quality. This is the first direct proof that the very large-sized transferred graphene layer obtained from epitaxial graphene on the SiC substrate by using the proposed transfer process had almost the same quality as the small-sized graphene transferred from the HOPG bulk. Each sub-peak showed good agreement with the theoretical predictions, because the red shift of the spectrum was due to the low-energy excitation. [19]

Figure 5 shows the photo-images of the graphene transferred on the $SiO_2$/Si substrate after the second stage in the transfer process. After the second stage, we can observe the graphene related Raman spectra from both the $SiO_2$/Si and the remaining Ti surface. Both Raman peaks show the sharp G- and G'-peaks came from mono-layer graphene. The optical contrast difference of the transferred graphene on the $SiO_2$/Si surface observed by the optical microscope also clearly shows the mono-layer graphene. This means that the bi-layer graphene on the Ti layer can be successfully separated the transferred mono-layer graphene onto the $SiO_2$/Si surface and the remaining mono-layer graphene on the Ti layer at the second stage in the transfer process. The Raman spectra from the mono-layer graphene on the $SiO_2$ layer and from the $SiO_2$ layer are shown in Fig. 5, with the schematic representation of the sample and the photo image. In this study, from the uniformity of the Raman spectra and the optical microscopic image, we could conclude that the first successful transfer of the mono-layer graphene (area: 2.5 mm×4 mm), from the bi-layer graphene seed layer on the Ti layer was achieved. Moreover, we could obtain the bi-layer graphene also from the tri-layer graphene seed layer on the Ti intermediate layer by using the proposed transfer process. However, considerably strong D-band peaks suddenly appeared in the Raman spectra from the mono-layer (or the bi-layer also) graphene on the $SiO_2$ layer although the sharp G- and the G`- band peaks due to the mono-layer (or the bi-layer also) graphene were observed clearly in the Raman spectra, as shown in Fig. 5. We think that there may be two possible causes of the enhancement in the D-peak: (1) the mechanical defect formation due to the exfoliation process at the second stage and (2) the large rippled structure of the transferred monolayer graphene due to the weak bonding onto the $SiO_2$ surface. We have to take into account the breaking of some of the Raman selection rules in the scattering process in the latter case. In fact, we could observe the graphene related Raman spectra from the surface with very similar contrast to the $SiO_2$ surface in the optical microscopic image. That is, the mono-layer graphene might have come in contact with the $SiO_2$ surface in a rippled, and not a uniform, manner. In such a case, we might be able to note down observations for an extra mode 一 a mode that is usually forbidden under the ordinary conditions, such as the D-band peak mode. However, we can not exclude a possible origin of the Raman peaks from $SiO_2$ surface due to the very tiny transferred graphene domains at the present stage. The most possible cause of the D-band appearance during the defect formation is now under investigation. If it is the case, maybe we need to develop some repair techniques for the defects of the graphene like annealing under the carbon related ambient in the future.

**Conclusion**

In this study, we have proposed a new transfer process; this is the first time that a large-area monolayer epitaxial graphene has been successfully transferred onto a $SiO_2$/Si substrate. That is, we found that the Ti intermediate layer formed by the electron beam evaporation process is very effective in exfoliating epitaxial graphene on the vicinal Si-terminated SiC substrate and in transferring graphene from a Ti intermediate layer onto a $SiO_2$ surface. We also observed that the monolayer and the bi-layer graphene were transferred onto the $SiO_2$ surface from the bi-layer and the tri-layer epitaxial graphene on the SiC substrates, respectively. Further, there might be no size limitation on the graphene transferred using the proposed process. Our results can open the door to a new field of graphene research.

**Acknowledgements**

This work was partly supported by KAKENHI (Contract No. 19360141 & 21246006) from

**Figures**

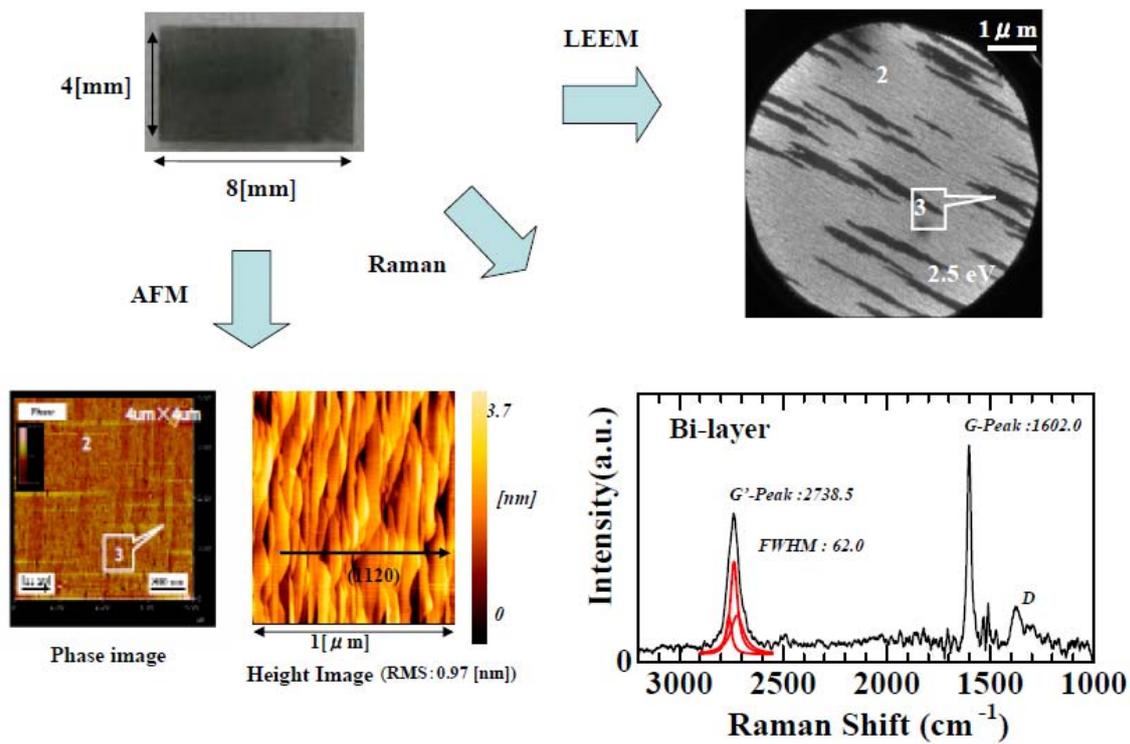

Fig. 1: Typical epitaxial graphene formed on vicinal Si-terminated SiC substrate (size: 4mm × 8mm). Pristine epitaxial graphene on the vicinal Si-face SiC substrate almost had the thickness of a bi-layer and a good layer quality as a result of high-temperature annealing. We found that the LEEM image is just corresponded to the AFM phase image, not height one.

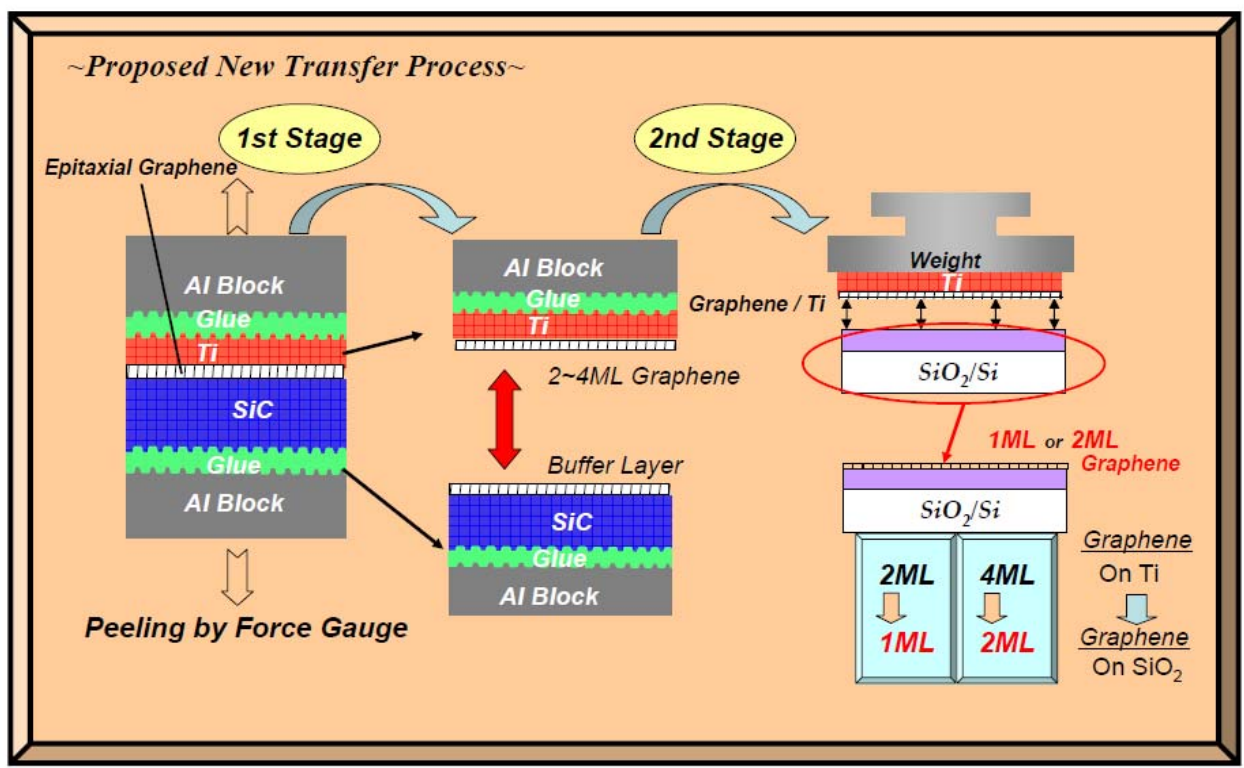

Fig. 2

Fig. 2: Schematic representation of proposed transfer process. We found that the Ti layer was very effective in transferring epitaxial graphene onto the SiC surface because of sufficient contact with the surface of epitaxial graphene at the first stage. We also found that the transfer of graphene was relatively easy from the Ti intermediate layer to the $SiO_2$/Si substrate. Furthermore, we could transfer the monolayer and/or the bi-layer graphene from the bi-layer and/or the tri-layer graphene seed on the Ti intermediate layer, respectively.

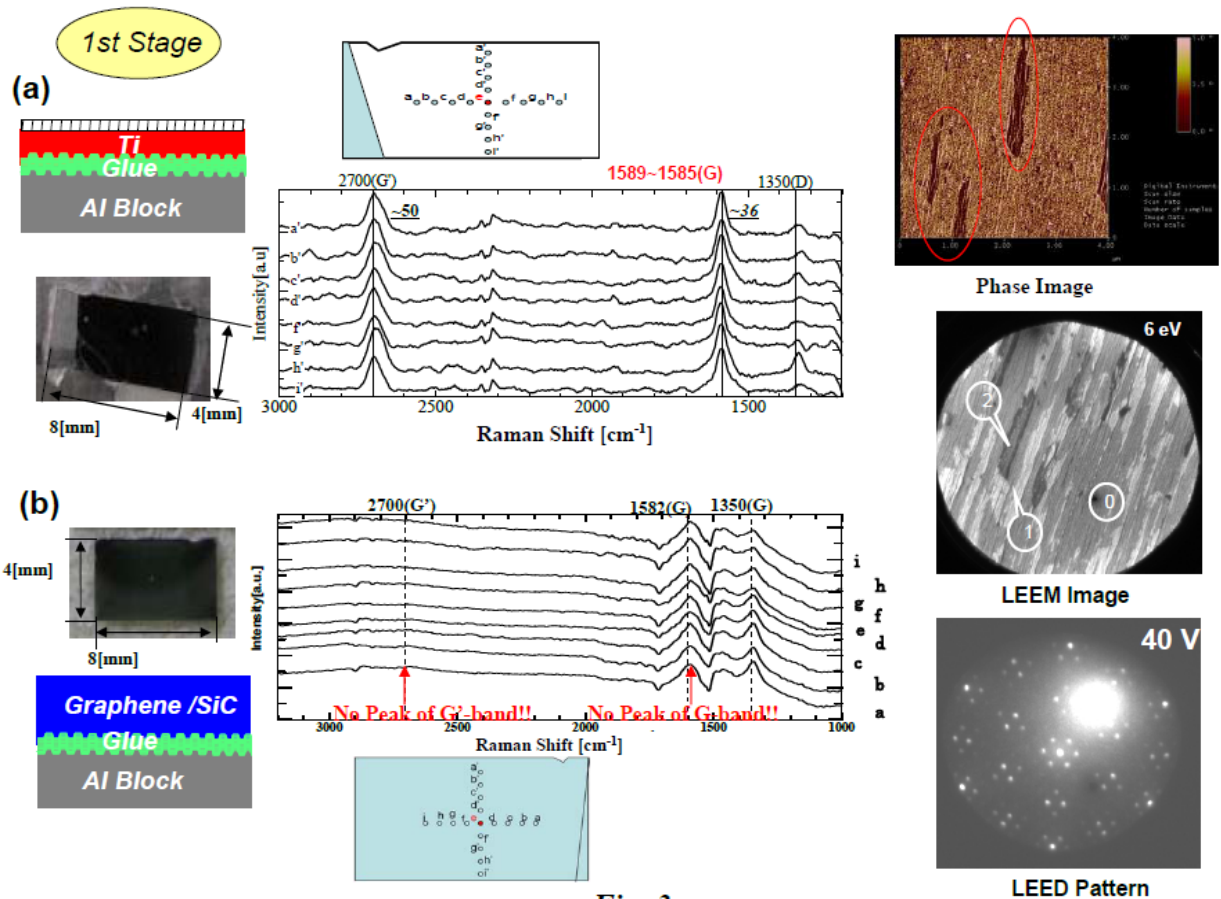

Fig. 3: Photo images, schematic representations of the samples, AFM phase image of graphene on Ti layer, LEEM image and LEED pattern from remaining SiC surface, and the Raman spectra from both samples of graphene transferred onto the Ti layer and/or the SiC substrate after the first stage in the transferred process. We could be confirmed successfully transfer the bi-layer graphene onto the Ti intermediate layer from SiC surface with the surface reconstruction of $6\sqrt{3} \times 6\sqrt{3}$ R30. The uniformity of the transferred graphene could be confirmed by the Raman measurement: we found no degradation in the quality of the transferred graphene at the first stage of the new transfer process.

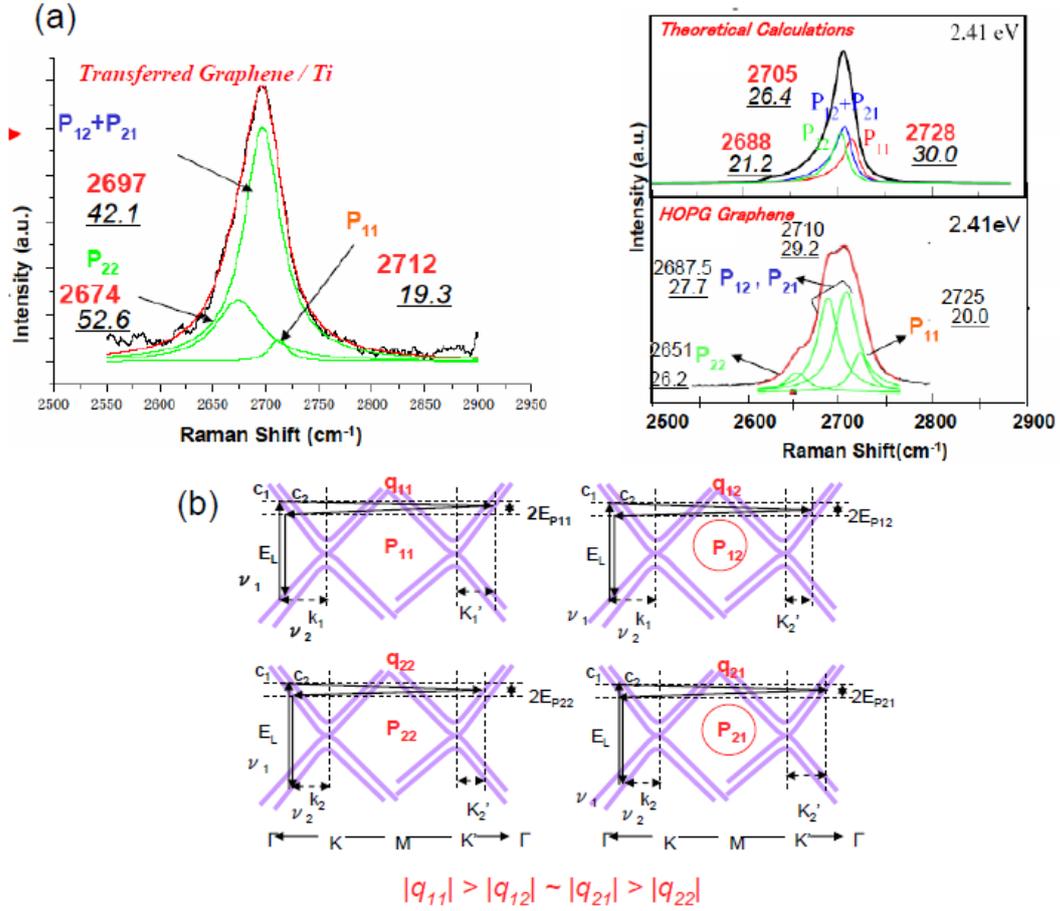

Fig. 4: Second-order Raman mode analysis of the graphene transferred onto the Ti layer in comparison with the theoretical calculation [22] and the experimental result from the small area transferred graphene in Ref. [23]. The second-order Raman spectrum was decomposed by the four processes, $P_{11}$, $P_{12}$, $P_{21}$ and $P_{22}$ as shown in Fig. 4 (b). The phonon momentum transfer in $P_{12}$ and $P_{21}$ was almost equal: therefore, the second-order Raman peak was decomposed by the three peaks of $P_{11}$, $P_{12}+P_{21}$ and $P_{22}$.

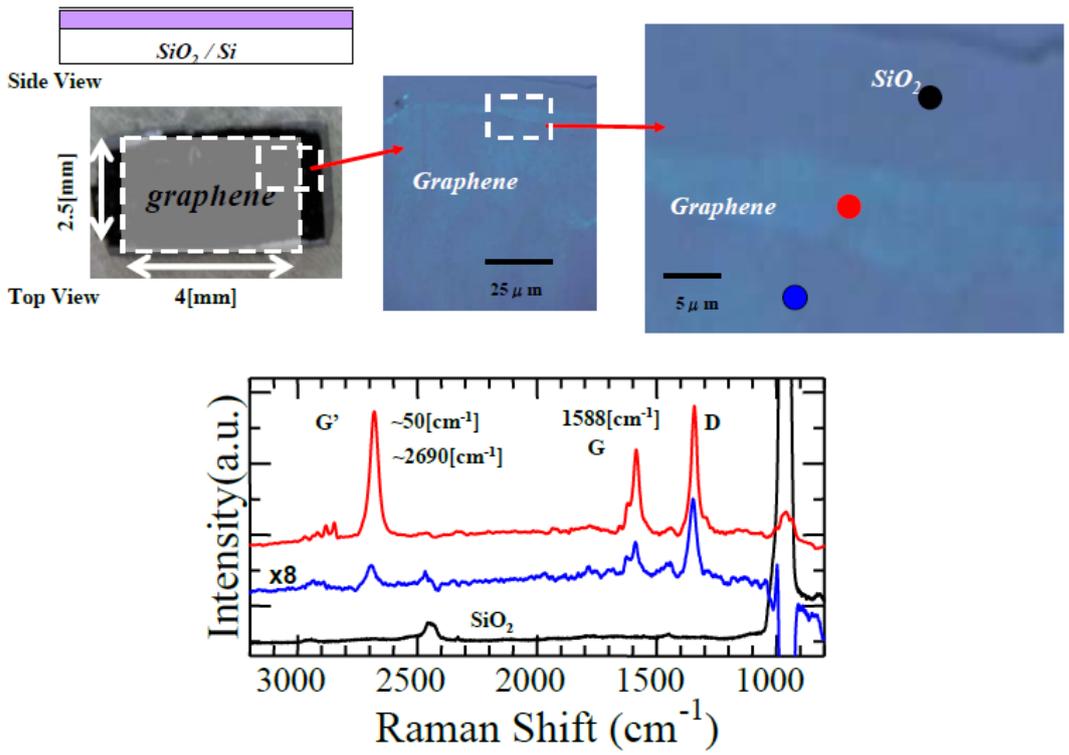

Fig. 5: Photo images, schematic representations, optical microscopic images of sample, and Raman spectra of graphene transferred onto SiO$_2$/Si substrate after the second stage of the transfer process. We could conclude that the first successful transfer of the monolayer (or the bi-layer) graphene with an area of 2.5 mm × 4 mm occurred from the bi-layer (or tri-layer) graphene seed layer onto the Ti intermediate layer. However, considerably strong D-band peaks suddenly appeared in the Raman spectra from the monolayer graphene on the SiO$_2$ layer although the sharp G-band and the G`-band peaks due to the monolayer graphene were observed clearly in the Raman spectra.